\newif\iffull\fullfalse
\newif\ifremarks\remarksfalse
\remarkstrue %makes REMARKs visible
\documentclass[runningheads,a4paper, oribibl]{llncs}

\usepackage{amssymb}
\usepackage{graphicx}

\usepackage{url}
\usepackage[para]{footmisc}
\usepackage{multirow}
\usepackage[table]{xcolor}
\usepackage{tabularx}
\usepackage{listings}
\usepackage{wrapfig}

\urldef{\mailsa}\path|{doliere.some, tamara.rezk, nataliia.bielova}@inria.fr|

\def\code#1{\texttt {\small{#1}}}
\def\it#1{\textit {{#1}}}

\newcolumntype{Y}{>{\centering\arraybackslash}X}

\def\REMARK#1#2{        % 1 = reviewer (1, 2, ...); 2 = comment
  \ifremarks
  \begin{center}
  \noindent\fbox{
  \begin{minipage}[b]{0.9\textwidth}
  \textbf{[#1]: #2}
  \end{minipage}}
  \end{center}
  \fi}

\def\NB#1{\REMARK{Nataliia}{#1}}

\def\SHORTEN{\vspace*{-0.5cm}}

\def\MPS{Middle Party Server}
\def\RS{Rewrite Server}

\def\incontext{in-context}
\def\crosscontext{cross-context}

\usepackage[utf8]{inputenc}
\usepackage{color}
\definecolor{lightgray}{rgb}{0.95, 0.95, 0.95}
\definecolor{darkgray}{rgb}{0.4, 0.4, 0.4}
\definecolor{purple}{rgb}{0.65, 0.12, 0.82}
\definecolor{editorGray}{rgb}{0.95, 0.95, 0.95}
\definecolor{editorOcher}{rgb}{1, 0.5, 0} % #FF7F00 -> rgb(239, 169, 0)
\definecolor{editorGreen}{rgb}{0, 0.5, 0} % #007C00 -> rgb(0, 124, 0)
\usepackage{listings}

\lstdefinelanguage{CSS}{
  keywords={color,background-image:,margin,padding,font,weight,display,position,top,left,right,bottom,list,style,border,size,white,space,min,width, transition:, transform:, transition-property, transition-duration, transition-timing-function},	
  sensitive=true,
  morecomment=[l]{//},
  morecomment=[s]{/*}{*/},
  morestring=[b]',
  morestring=[b]",
  alsoletter={:},
  alsodigit={-}
}

\lstdefinelanguage{CSP}{
  keywords={default-src, script-src, child-src, frame-src, script-src, style-src, report-uri, connect-src, img-src, object-src, frame-ancestors, plugin-types, form-action, sandbox, worker-src, font-src, media-src},
  morestring=[b]',
  alsoletter={:},
  alsodigit={-}
}

\lstdefinelanguage{JavaScript}{
  morekeywords={typeof, new, true, false, catch, function, return, null, catch, switch, var, if, in, while, do, else, case, break},
  morecomment=[s]{/*}{*/},
  morecomment=[l]//,
  morestring=[b]",
  morestring=[b]'
}

\lstdefinelanguage{HTML5}{
  language=html,
  sensitive=true,	
  alsoletter={<>=-},	
  morecomment=[s]{<!-}{-->},
  tag=[s],
  otherkeywords={
  >,
	<!DOCTYPE,
  </html, <html, <head, <title, </title, <style, </style, <link, </head, <meta, />,
	</body, <body,
	</div, <div, </div>, 
	</p, <p, </p>,
	</script, <script,
  <canvas, /canvas>, <svg, <rect, <animateTransform, </rect>, </svg>, <video, <source, <iframe, </iframe>, </video>, <image, </image>
  },
  ndkeywords={
  =,
  charset=, src=, id=, width=, height=, style=, type=, rel=, href=,
  fill=, attributeName=, begin=, dur=, from=, to=, poster=, controls=, x=, y=, repeatCount=, xlink:href=,
  margin:, padding:, background-image:, border:, top:, left:, position:, width:, height:,
  transform:, -moz-transform:, -webkit-transform:,
  animation:, -webkit-animation:,
  transition:,  transition-duration:, transition-property:, transition-timing-function:,
  }
}

\def\SHORTEN{\vspace*{-0.5cm}}

\begin{document}

%\title{Towards The Design and Implementation of Privacy-Preserving Web Applications}
\title{Control What You Include!}
\subtitle{Server-Side Protection against Third Party Web Tracking}
\titlerunning{Server-Side Protection against Third Party Web Tracking}
%\title{Automatic Server-Side Protection against Third-party Web Tracking}
%\titlerunning{Privacy-Preserving Web Applications}

\author{Doli\`ere Francis Som\'e\and Nataliia Bielova\and Tamara Rezk}
\institute{Universit\'e C\^ote d'Azur, Inria\\
\mailsa%\\
%\url{https://www.inria.fr}
}

\maketitle

\begin{abstract}
%Third party tracking is a mechanism for surveilling users on the Web. 
%Website developers often need to include third party 
%content in order to provide basic functionality of the website: 
%recent studies show that % 95\% of websites contain at least 
Third party tracking is %a mechanism for surveilling users on the Web. 
the practice by which third parties recognize users accross different websites 
as they browse the web.
%Website developers often need to include third party 
%content in order to provide basic functionality.  %of the website: 
Recent studies show that % 95\% of websites contain at least 
%one third party content, out of which 88\% are tracking the users. 
90\% of websites contain third party content that is tracking 
its users across the web. 
%A great advantage of including the third-party content is 
%in the monetisation by advertisements, the use of already developed libraries, and 
%building mashups (such as websites with maps), where a website can interact with the third-party content.
%%
%However, when a website developer includes a third party content, 
%the developer cannot know whether the third party contains tracking mechanisms.
%If a website developer wants to protect its users from being tracked, 
Website developers often need to include third party 
content in order to provide basic functionality.  %of the website: 
However, when a developer includes a third party content, 
she cannot know whether the third party contains tracking mechanisms.
If a website developer wants to protect her users from being tracked, 
%she has no means to do it in the current model of the Web. 
%In order to keep the 
%promise of non-tracking, 
the only solution is to exclude any third-party content, thus trading functionality for privacy. 
%Moreover, the users do not trust the website 
%owners, and must protect their privacy by installing client-side protection mechanisms. 
%(AdBlock Plus, Ghostery, Privacy Badger).

We %propose a tool 
describe and implement a privacy-preserving web architecture
that gives website developers a control over third party tracking: 
%Using our tool, website owners 
 developers are able to include functionally useful third party content, the same time ensuring that the end users are not tracked by the third parties. 
\end{abstract}

\section{Introduction}

%- Web tracking is everywhere

%Third party tracking is a well-known mechanism for surveilling users on the Web. 
%In recent years, tracking technologies have been extensively studied and measured~\cite{Engl-Nara-16-CCS, Kris-Will-09-WWW, Maye-Mitc-12-SP, Roes-etal-12-NSDI, Eckersley-10-PETS, Niki-etal-13-SP} -- researchers have found that third parties embedded in websites use numerous technologies, such as  third-party cookies, HTML5 local storage, browser cache and others to store identifiers that can be used to recognize users across websites~\cite{Solt-etal-10-AAAI} and build browsing history profiles.  
%Third party tracking is a well-known mechanism for surveilling users on the Web. 
Third party tracking is %a mechanism for surveilling users on the Web. 
the practice by which third parties recognize users accross different websites 
as they browser the web.
In recent years, tracking technologies have been extensively studied and measured~\cite{Engl-Nara-16-CCS, Kris-Will-09-WWW, Maye-Mitc-12-SP, Roes-etal-12-NSDI, Eckersley-10-PETS, Niki-etal-13-SP} -- researchers have found that third parties embedded in websites use numerous technologies, such as  third-party cookies, HTML5 local storage, browser cache and 
device fingerprinting
%others %to store identifiers that can be 
that allow the third party to recognize users across websites~\cite{Solt-etal-10-AAAI} and build browsing history profiles.  
Researchers found that more than 90\% of Alexa top 500 websites~\cite{Roes-etal-12-NSDI} contain 
third party web tracking content, while some sites include as much as 34 distinct third party contents~\cite{Lern-etal-16-USENIX}.

%Even though the tracking privacy threat involves three parties -- first parties (main websites), 
%web users and third parties -- current studies on web tracking measurement have exclusively 
%focused only on the web users and third parties. 
But why do website developers include so many third party contents 
(that may track their users)? %in their applications?
%Third party content is an essential part of the modern web applications. 
% SHORTER VERSION:
%Third party contents help %enhance user experience, 
%ease web application development with JavaScript libraries;
%enhance user experience with social widgets or other 
%third party components for mashups, such as maps; 
%include advertisements for the financial support of the website;
% and provide analytics data to websites owners.
Though some third party content, such as images and CSS~\cite{CSS} files can be copied to the main 
(first-party) site, such an approach has a number of disadvantages for other kinds of content.
%\begin{itemize}
%\item 
\emph{Advertisement} is the base of the economic model in the web -- without advertisements 
many website providers will not be able to financially support their website maintenance.
%
%\item 
\emph{Third party JavaScript libraries} offer extra functionality: though copies of 
such libraries can be stored on the main first  party site, this solution will sacrifice 
maintenance of these libraries when new versions are released. The developer 
would need to manually check the new versions.
%
%\item 
\emph{Web mashups}, as for example applications that use hotel searching together 
with maps, are actually based on reusing third-party content, as well as maps, and would not be able to provide their basic functionality without including the  third-party content.  
%\end{itemize}

%Therefore, including 
Including JavaScript libraries, content for mashups or advertisements means 
that the web developers cannot provide %promise 
to the users %that the web application will not 
%track its users. 
the guarantee of non-tracking.
Hence, the promise to provide privacy %to end users 
has  a very high cost because there 
are no existing automatic tools to maintain control of third party tracking % privacy 
%of the users of 
on the website. 
%Therefore, to 
To keep a promise of non-tracking, % to end users, 
the only solution today 
is to exclude any third-party content\footnote{For example, see \url{https://duckduckgo.com/}.}, 
thus trading functionality for privacy. 

%Without sacrificing financial support, maintenance, and functionality: 
%\emph{What can we propose to provide Web developers the ability 
%to keep a promise of non-tracking to their end users? }

In this paper, we present a new Web application architecture that 
allows web developers to gain control over certain types of third party content. 
Our solution is based on the automatic rewriting of the web application in such 
a way that the third party requests are redirected to a trusted web server, 
with a different domain than the main site. 
This trusted web server may be either controlled by a trusted party, 
or by a main site owner -- it is enough that the trusted web server has a different domain. 
A trusted server is needed so that the user's browser will treat all redirected requests 
as third party requests, like in the original web application.
The trusted server automatically eliminates third-party tracking 
cookies and other technologies.
%as well as adds new JavaScript code to the third party 
%libraries that would ensure that no cookies and other storages
%are stolen by the third party library. 
%Moreover third party contents are important to first party web sites owners. However, since any third party content, even benign ones, can participate in tracking, we discuss the challenges and opportunities
%of designing and implementing privacy-preserving websites. 
%

%While analysing the applicability of our idea to the existing Web application, 
%we have discovered that not every third party content can be controlled by a web developer. 
%For example, \code{iframe} elements that originate from a different domain 
%cannot be fully controlled by a first party website, though we can mitigate 
%their tracking capabilities (see Section~\ref{sec:solution}). 
%We therefore make the following contributions:
In summary our contributions are:
\begin{itemize}
%\item We first provide a classification of third party contents from a first party perspective. Depending on the type of third party contents and the way they are embedded in a website, 
%they either execute in a browsing context controlled by the first party 
%developer, or in a different context over which the developer has no control. 
\item A classification of third party contents that can and cannot 
be controlled by the website developer. 
\item An analysis of third party tracking capabilities -- we analyse two mechanisms: 
recognition of a web user, and identification of the website she is visiting
% she is interacting with
\footnote{Tracking 
is often defined as the ability of a third party to recognize a user through different websites. 
However, being able to identify the websites a user is interacting with is equally crucial 
for the effectiveness of tracking.}.
\item %We propose a 
A new architecture that allows to include third party 
content in web applications and eliminate stateful tracking. 
\item %We provide an 
An implementation of our architecture, demonstrating its effectiveness 
at preventing stateful third party tracking in several websites.
\end{itemize}

%\section{Background and Threat Model}
\section{Background and Motivation}
\label{sec:background}

%\subsection{Classification of Third Party Tracking}
%\label{sec:classification}

%%%%%% stateful TRACKING %%%%%%
Third party web tracking is the ability of a third party 
to re-identify users as they browse the web and record their browsing history~\cite{Maye-Mitc-12-SP}. 
Tracking is often done with the purpose of web analytics, targeted advertisement, 
or other forms of personalization. 
The more a third party is prevalent among the websites a user interacts with, 
%closer 
the more precise is %will be 
the %built 
browsing history collected by the tracker. %, to the real browsing history of the user.
Tracking has often been conceived  as the ability of a third party 
to recognize the web user.
However, %building a browsing history require two conditions to be met, when a request is received by the third party server:
for successful tracking, each user request should contain two components:
%\begin{itemize}
% \item Recognize the user making the request.
% \item Identify the website from which the request is being made.
%\end{itemize}
%Thus, tracking information include both website identification and user recognition information.
\begin{description}
\item[User recognition] is the information that allows tracker to recognize the user;
%(either an identifier stored on the user's machine, or a device fingerprint);
\item[Website identification] is the website which the user is visiting.
\end{description}

%%%%%%%%% EXAMPLE %%%%%%%%

% ADD FIGURE

\begin{figure}[!tp]
\centering
    \includegraphics[width=0.5\textwidth]{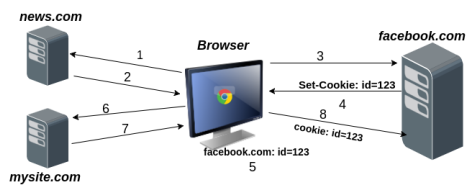}
    \caption{Third Party Tracking}
    \label{fig:tracking}
    \SHORTEN
\end{figure}

For example, when a user visits \code{news.com}, the browser may make 
additional requests to \code{facebook.com}, as a result, Facebook learns 
about the user's visit to \code{news.com}. Figure~\ref{fig:tracking} shows 
a hypothetical example of such tracking where \code{facebook.com} is the third party.

Consider that a third party server, such as \code{facebook.com} hosts 
different contents, and some of them are useful for the website developers.
The web developer of another website, say \code{mysite.com}, 
 would like to include such \emph{functional} 
content from Facebook, such as Facebook "Like" button, an image, 
or a useful JavaScript library, 
but the developer does not want its users to be tracked by Facebook. 
If the web developer simply includes third party Facebook content 
in his application, all its users are likely to be tracked by 
cookie-based tracking.
%Suppose that a user interacts with a website \code{site.com} that embeds 
%\emph{tracking} content from \code{third.com}, as a result, \code{third.com} 
%stores a unique identifier in the third party cookies of the user's browser.
%
%The user then interacts with another website \code{example.com} that embeds 
%a \emph{functional} content from \code{third.com}, then the previously set 
%unique identifier, stored as a cookie, is attached by the browser to all the 
%requests to that third party\footnote{If the two contents are originating from 
%the same domains and paths}. 
Notice that each request to \code{facebook.com} also contains an 
HTTP Referrer header, automatically attached by the  browser. 
This header contains the website URL that the user is visiting, which 
 allows Facebook to build user's browsing history profile. 
 
 The example demonstrates cookie-based tracking, which is extremely 
 common~\cite{Roes-etal-12-NSDI}. Other types of third party tracking, 
 that use other client-side storage mechanisms, such as HTML5 LocalStorage, 
 or cache, and device fingerprinting that do not require any storage 
 capabilities, are also becoming more popular~\cite{Engl-Nara-16-CCS}.
 
%%%%%%%%%%%%%%%%%%%%%%%%%%
\subsubsection{Web developer perspective}

A web developer may include third party content in her webpages, 
either because this content \emph{intentionally} tracks users 
(for example, for targeted advertising), or because this content 
is important for the functioning of the web application. 
We therefore distinguish two kinds of third party contents from 
a web developer perspective: \emph{tracking} and \emph{functional}. 
\emph{Tracking} content is \emph{intentionally} embedded by website owner 
for tracking purposes. \emph{Functional} content is embedded in a webpage
for other purposes than tracking: for example, JavaScript libraries that provide
additional functionality, such as jQuery, or other components, such as maps.
%not for tracking purposes. However, they can \emph{accidentaly} participate in tracking. 
%
In this work, we focus on \emph{functional} content and investigate the 
following questions: 
\begin{itemize}
\item What kind of third content is possible to control from a server-side (web 
developer) perspective?
\item How to eliminate the two components of tracking (user recognition 
and website identification) from the functional third party that the website 
embeds?
\end{itemize}

%Notice that functional content may be tracking the users even though  it was 
%not the intention of the web developer. 
%%To illustrate this, let's 
%Consider a third party server \code{third.com} that hosts both 
%{tracking} and {functional} contents.
%Suppose that a user interacts with a website \code{site.com} that embeds 
%\emph{tracking} content from \code{third.com}, as a result, \code{third.com} 
%stores a unique identifier in the third party cookies of the user's browser.
%%
%The user then interacts with another website \code{example.com} that embeds 
%a \emph{functional} content from \code{third.com}, then the previously set 
%unique identifier, stored as a cookie, is attached by the browser to all the 
%requests to that third party\footnote{If the two contents are originating from 
%the same domains and paths}. 
%Each request to \code{third.com} also contains an HTTP Referrer header, 
%automatically attached by the  browser, that allows the \code{third.com} 
%to identify the website a user visits (in our case, \code{example.com}.
%With website identification sent along with the request, the third party can augment the user browsing profile with the current website. 

%%% BROWSING CONTEXT

\subsection{Browsing Context}
\label{sec:context}

%\subsubsection{Same Origin Policy}
Browsers implement different specifications to securely fetch and aggregate third party content. 
One widely used approach is the the \emph{Same Origin Policy (SOP)}~\cite{Same-Origin-Policy}, 
a security mechanism designed for developers to 
isolate legacy content from potentially untrusted third party content. An origin is defined as scheme, host and port number, of the URL\footnote{\url{https://www.w3.org/TR/url/}} of the third party content.
%SOP applies only to third party content that is executed in a different browsing 
%context %(Section~\ref{sec:context}) 
%than the website that embeds this content.
%Basically all third party contents have a different origin from that of the first party webpage in which they are embedded. Third party contents served by the same third party, share the same origin.

%\subsubsection{Browsing Context}
When a browser renders a webpage delivered by a first party, 
the page is placed within a \emph{browsing context}~\cite{Context}. 
A browsing context represents an instance of the browser in which a document such as a webpage is displayed to a user, for instance browser tabs, and popup windows. 
Each browsing context contains 1) a copy of the browser properties (such as 
browser name, version, device screen etc), stored in a specific object;
%Properties of the browser are duplicated in each context, for instance, objects describing the browser characteristics such as its name, version, the device screen size etc.
2) other objects that depend on the origin of the document according to SOP.
% displayed in the context. 
%On the contrary, the values of other objects depend on the origin of the document displayed in the context. 
For instance, the object \code{document.cookie} gives the cookies 
related to the origin of the current context.
\\

{\bf In-context and cross-context content}
Certain types of content embedded in a webpage, such as images, links, and scripts, 
are associated with the context of the webpage, and we call them \emph{\incontext} content.
Other  types of content, such as \code{<iframe>}, \code{<embed>}, and \code{<object>} tags 
are associated with their own browsing context, and we call them \emph{\crosscontext} content. 
% - we refer to them as \incontext\ contents. That is of the exception of contents embedded using one of the \code{<iframe>, <embed>, <object>, <frame>, <frameset>, <applet>} HTML tags.
%We refer to them as \crosscontext contents. The latters are associated with their own browsing contexts, called nested browsing contexts. 
Usually, \crosscontext\ content, such as \code{<iframe>} elements, 
cannot be visually distinguished from the webpage 
in which they are embedded, however they are as autonomous as other browsing contexts, 
such as tabs or windows.
%, 
%because they are 
%displayed with other contents as a single unit. But, they are as autonomous as other browsing contexts. 
Table~\ref{tab:third_party_content} shows different third party contents and their execution contexts.

\begin{table}[!ht]
\centering
\begin{tabular}{cc|c|c}
\cline{2-3} & \multicolumn{1}{ |c| }{HTML Tags} & Third Party Content\\ 
\cline{1-3}
%\multicolumn{1}{ |c  }{\multirow{5}{*}{\incontext} } &
\multicolumn{1}{ |c  }{\multirow{5}{*}{{\bf \incontext}} } &
\multicolumn{1}{ |l| }{\code{<link>}} & Stylesheets\\ 
\cline{2-3}
\multicolumn{1}{ |c  }{} & \multicolumn{1}{ |l| }{\code{<img>}} & Images\\ 
\cline{2-3}
\multicolumn{1}{ |c  }{} & \multicolumn{1}{ |l| }{\code{<audio>}} & Audios\\ 
\cline{2-3}
\multicolumn{1}{ |c  }{} & \multicolumn{1}{ |l| }{\code{<video>}} & Videos \\ 
\cline{2-3}
\multicolumn{1}{ |c  }{} & \multicolumn{1}{ |l| }{\code{<form>}} & Forms \\ 
\cline{2-3}
\multicolumn{1}{ |c  }{} & \multicolumn{1}{ |l| }{\code{<script>}} & Scripts \\ 
\cline{1-3}
%\multicolumn{1}{ |c  }{\multirow{2}{*}{\crosscontext} } & 
\multicolumn{1}{ |c  }{\multirow{2}{*}{\bf cross-context} } & 
\multicolumn{1}{ |l| }{\code{<(i)frame>, <frameset>, <a>}} & Web pages \\ 
\cline{2-3}
\multicolumn{1}{ |c  }{} & \multicolumn{1}{ |l| }{\code{<object>, <embed>, <applet>}} & Plugins and Web pages \\
\cline{1-3}
\\
\end{tabular}
\caption{Third party content and execution context.}
\label{tab:third_party_content}
\SHORTEN
\end{table}

The Same Origin Policy manages the interactions between different browsing contexts. In particular, it prevents \incontext\ scripts from interacting with the content from a \crosscontext\ content in case their origins are different.
To communicate, both contexts should rely on inter-frame communications APIs such as \code{postMessage}~\cite{PostMessage}.

\subsection{Third Party Tracking}

In this work, we consider only stateful tracking technologies -- they require an identifier be stored client-side, the most common storage mechanism is cookies, but others, such 
as HTML5 LocalStorage and browser cache are also stateful tracking mechanisms. 
Figure~\ref{tab:stateful_tracking_mechanisms} %summarizes the related stateful tracking information.
presents the well-known stateful tracking mechanisms. We distinguish two components 
necessary for successful tracking: user recognition and website identification. 
For each component, we describe the capabilities of \incontext\ and \crosscontext. 
We also distinguish \emph{passive tracking} (done through HTTP headers) and \emph{active tracking} (through JavaScript or plugin script execution).

\begin{figure}[!ht]
\centering
\begin{tabular}{|l|l|l|l|l}
\cline{2-5}
\multicolumn{1}{c}{} & 
\multicolumn{2}{|c|}{\bf User Recognition} &\multicolumn{2}{|c|}{\bf Website Identification} %& \multicolumn{1}{|c}{}
\\
\cline{2-5}
\multicolumn{1}{c|}{} &
%\begin{tabular}{c}Passive \end{tabular}  
\multicolumn{1}{c|}{Passive}
& \multicolumn{1}{c|}{Active} 
& \multicolumn{1}{c|}{Passive}
& \multicolumn{1}{c|}{Active} \\ 
\hline
 \multicolumn{1}{ |c|  }{\rotatebox{60}{\bf \incontext}}  &
\multirow{2}{*}{
\begin{tabular}{l} 
%\emph{cookie, set-cookie} \\ 
{HTTP cookies} \\ 
{Cache-Control} \\  
{Etag} \\  
%\emph{If-Modified-Since}  \\%\\
%\emph{Set-Cookie} \\
{Last-Modified} \\
%\emph{If-Modified-Since}
\end{tabular}
} & 
%
%\begin{tabular}{l}%Plugins Storages \\  \hspace{1em}
%{Flash LSOs} \end{tabular} 

\multicolumn{1}{c|}{-}& 
\begin{tabular}{l} {Referer} \\ {Origin}\end{tabular} &
 \multicolumn{1}{ c|  } {\begin{tabular}{l}\code{document.URL} \\  \code{document.location} \\ \code{window.location}\end{tabular}} 
%& \multicolumn{1}{ c|  }{\rotatebox{90}{\incontext}} 
\\
\cline{1-1}\cline{3-5}
%\hline
%
\multicolumn{1}{ |c|  }{\rotatebox{60}{\bf \crosscontext}} &
% empty 
& 
\begin{tabular}{l}{Flash LSOs} \\  {\code{document.cookie}} \\  {\code{window.localStorage}} \\ {\code{window.indexedDB}}\end{tabular} & 
{Referer} &
 \multicolumn{1}{ c|  } {\code{document.referrer} }
% & \multicolumn{1}{ c|  }{\rotatebox{90}{\crosscontext}} 
\\
\hline 
\end{tabular}
\caption{Stateful tracking mechanisms}
\label{tab:stateful_tracking_mechanisms}
\SHORTEN
\end{figure}

{\bf In-context tracking}
 In-context third party content is associated with the browsing context 
 of the webpage that embeds it (see Table~\ref{tab:third_party_content}). 

 \emph{Passively}, such content may use HTTP header to recognize the 
 user and identify the visited website. 
When a webpage is rendered, the browser sends a request to fetch all 
third party contents embedded in the page. The response from the third party 
with the requested content may contain HTTP headers that may be used for 
tracking. For example, \emph{Set-cookie} HTTP header tells the browser to 
save the third party cookies, that will be later automatically attached 
to every request to this third  party in the Cookie header. 
Etag HTTP header and other cache 
mechanisms like \emph{Last-Modified} and \emph{Cache-Control} HTTP 
headers may also be used to store user identifier~\cite{Solt-etal-10-AAAI}.
%That is when passive tracking information are automatically exchanged between the browser and the third party. 
%The main challenge here is that the first party websites owners do not control those interactions.
To identify the visited website, a third party can either check the 
\emph{Referer} HTTP header, automatically attached by the browser, 
or an \emph{Origin} header\footnote{Origin header is also automatically generated 
by the browser when the third party content is trying to access data 
using Cross-Origin Resource Sharing~\cite{CORS} mechanism.}.

{\em Actively}, \incontext\ third party content cannot use browser storage mechanisms, 
such as cookies or HTML5 Local Storage associated to the third party because of 
%Because of 
the limitations imposed by the SOP (see Section~\ref{sec:context}). %, an \incontext third party content cannot perform active user recognition. 
For example, if a third party script uses \code{document.cookie} API, 
%it is accessing the cookies on the current page, and not that of the related third party.
it is able only to read the cookies of the main website, but not those associated to the third party.
This allows tracking within the main website but does not allow tracking 
cross-sites~\cite{Roes-etal-12-NSDI}. 
%\NB{FRancis to add Flash LSOs description}
%
For website identification, third party active content, such as JavaScript, 
can use several APIs, such as \code{document.location} and others. %as discussed in Table~\ref{tab:stateful_tracking_mechanisms}.
%The challenge here is to prevent access to those APIs that help identify the website.
%Nonetheless, since those contents executes within a context which is controlled by the first party website owner, the latter can implement access control over those APIs.

{\bf Cross-context tracking} 
Cross-context third party content, such as \code{iframe}, is associated
with the browsing context of the third party 
%rather than with then webpage that embeds it 
that provided this content. 
%(see Table~\ref{tab:third_party_content}). 

{\em Passively,} the browser may transmit HTTP headers used for user recognition 
and website identification, just like with the
%passive tracking information to fetch \crosscontext third party contents. 
\incontext\ third party content. %Similary to \incontext contents, the request to fetch a \crosscontext third party content will contain the URL of the embedding webpage. 
Every third-party request for \crosscontext\ content 
will contain the URL of the embedding webpage in its \emph{Referer} header.
Note that this is true only for the \crosscontext\ content, say an \code{<iframe>}, directly embedded in the webpage.
Within the iframe, there may be additional third party contents. Since they are not  embedded directly in the webpage, and because the iframe is an autonomous though nested browsing context,
requests to fetch contents embedded within this context will carry, not the URL of the webpage, but that of the iframe in their \emph{Referer} header, and the origin of the iframe in their CORS requests \emph{Origin} header. 
%However, if the \crosscontext contains additional third party contents embedded within it, the \emph{Referer} header in their request headers will contain the URL of 
%the \crosscontext, and not that of the page in which the \crosscontext is embedded. 
%The \emph{Origin} header will not provide a website identification because 
%when a 
%Similarly, if 
%Cross-Origin Resource Sharing~\cite{CORS} request is made from the 
%\crosscontext, the browser will set an \emph{Origin} header to the URL 
%of the  third party that issued the request.
%Its value is also the origin of the \crosscontext, and not the that of the webpage in which the \crosscontext is embedded.

%In case of passive tracking, the only 
%header that is not transmitted for \crosscontext\ content, is Origin header.
%\NB{Francis, explain why Origin header is not possible in this case}
%
%It is worth noting the passive website identification carried by those requests is the URL of the \crosscontext, and not the URL of the parent \incontext. 
%\NB{I don't understand the meaning of the sentence above. Francis, try to rephrase}

{\em Actively}, \crosscontext\ third party content can use a number of APIs to 
store user identifier in the browser. These APIs include  cookies 
(\code{document.cookie}), HTML5 LocalStorage (\code{document.localStorage}), 
IndexedDB, and Flash Local Stored Objects (LSOs). 
For  website identification,  \code{document.referrer} API can be used -- it 
 returns the value of HTTP Referrer header transmitted to the third party 
when the third party content was fetched.
%Note that this API holds the same passive website information that has been transmitted to the third party when fetching the \crosscontext content.
%So, if no passive website identification were transmitted to the third party during the request, this API holds no information about the website.
%
Because \crosscontext\ third party is associated with its own browsing context, 
it is able to embed even more third party contents within this \crosscontext. 
%
%Within a \crosscontext, a third party has a full spectrum of user recognition active APIs (See Table~\ref{tab:stateful_tracking_mechanisms}.
%This is especially challenging from a developer perspective.

{\bf Combining \incontext\ and \crosscontext\ tracking}
Imagine a third party script from \code{third.com} embedded in a webpage -- according to the 
context and to the SOP, it is \incontext. If the same webpage embeds another 
third party content from \code{third.com}, which is \crosscontext, 
then because of SOP, such script and iframe cannot interact directly.
However, script and iframe can still communicate through 
inter-frame communication APIs such \code{postMessage}~\cite{PostMessage}. 

This communication between different contexts allow them to exchange the 
user identifiers and the website that the user visits. Efficient implementation 
of such combination of tracking may profit from easily 
implementable user recognition by \crosscontext\ code using, say \code{document.cookie}, 
and website identification by \incontext\ through various APIs such as \code{document.location}.
 %
%Even though the SOP prevents direct interactions between \incontext and \crosscontext content, two scripts running in both contexts and controlled by the same third party,
%can cooperate to exchange tracking information. The \crosscontext script recognizes the user while the \incontext script identifies the website. Moreover, the two scripts
%can exchange those information through the use of inter-frame communication APIs such \it{postMessage}~\cite{PostMessage} and later on transmit them to the related third party.
%This is a particularly prevalent configuration which is used by most websites to embed social widgets, advertisement, analytics scripts, etc. 
%Indeed, in order to embed a social widget for instance,
%a web developer first embeds a script from the third party social widget provider. 
For example, social widgets, such as Facebook "Like" button, or Google "+1" button, 
may be included in the webpages as a script. 
When the social widget script is executed on the client-side, it loads additional scripts, 
and new browsing contexts (iframes) %in which all the logic of the social widget is executed, 
allowing the third party to benefit from both \incontext\ and \crosscontext\ 
capabilities to track users.

\section{Privacy-preserving Web Architecture}
\label{sec:solution}

For third party tracking to be effective, it is necessary that it has two capabilities:
1) it is able to identify the website in which it is embedded, and 2) 
to recognize the user interacting with that website.
Disabling only one of these two capabilities for a given 
third party already prevents tracking. 
In order to mitigate the stateful tracking (see Section~\ref{sec:background}), 
we make the following design choices in our architecture:

\begin{enumerate}
 \item {\bf In-context content: prevent only user recognition}.
Preventing passive user recognition for \incontext\ content, such as 
images, forms and scripts is possible by removing HTTP headers such 
as \emph{Set-cookie, ETag} and others. 
%\NB{Francis, say how we prevent Flash LSOs}
However, it is particularly difficult to remove active website identification 
because trying to alter or redefine \code{document.location} and \code{window.location} APIs, 
will cause the main page to reload.%, which will be visible for the users. 
% This choice is based on the resilience of some active website identification mechanisms, 
%  such as \code{document.location} and \code{window.location}. The first party owner has more control over \incontext\ third party contents. Moreover, they cannot perform active stateful tracking.
%
 \item {\bf Cross-context content: prevent only website identification}. 
 %This other choice is based on the resilience user recognition within a \crosscontext. As we have already discussed, 
% Because of the limitations
% of the SOP, a website owner has no control over the \crosscontext\ third party 
% content, such as iframes. Therefore, it is not possible to modify the results 
% of storage access APIs, such as \code{document.cookie}. We discuss other 
% possibilities to block such APIs in Section~\ref{sec:discussion}.
 %We have already mentioned that if we present the 
 We prevent passive website identification %from being passively leaked to the third party, 
 by instructing the browser not to send HTTP \emph{Referer} header along with requests to fetch a \crosscontext\ content.
 Therefore, when the \crosscontext\ gets loaded, active website identification is impossible. Indeed, executing \code{document.referrer} returns not the URL of the embedding page, but an empty string.
 %this successfully prevents active website identification. % from within the website.
 %\NB{Francis, explain why document.referer is no longer useful}
 Because of the limitations
 of the SOP, a website owner has no control over the \crosscontext\ third party 
 content, such as iframes. Therefore, it is not possible to modify the results 
 of storage access APIs, such as \code{document.cookie}. We discuss other 
 possibilities to block such APIs in Section~\ref{sec:limitations}.
 \item {\bf Prevent  communication between \incontext\ and \crosscontext\  contents}. 
Our architecture proposes a way to block such communications that 
can be done by  \code{postMessage} API. We discuss the limitations of this 
approach in Section~\ref{sec:limitations}. 
\end{enumerate}

%In this section, we limit ourselves discuss stateful tracking based on \emph{functional} third party contents. We do not consider \emph{tracking} contents because they are intentionnaly embedded by first party websites owners
%to serve such purposes. Mitigating stateless tracking require advanced investigations. We leave this for future work.

\begin{figure}[t]
    \includegraphics[width=1\textwidth]{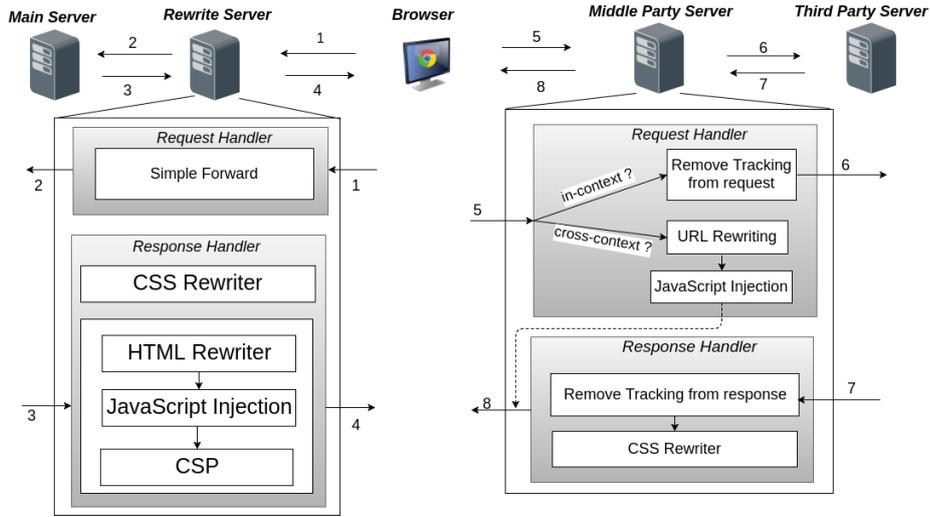}
    \caption{Privacy-Preserving Web Architecture}
    \label{fig:architecture}
    \end{figure}
% Figure~\ref{fig:architecture} presents the overall architecture we propose. 
% The main difference with traditional web application architecture is the introduction of two new components: a rewrite server and a middle party server.

%\section{Privacy-preserving Web Architecture}
%\label{sec:solution}

To help web developers keep their promises of non-tracking and still include 
third-party content in their web applications, we propose a new Web application 
architecture. % shown in  Figure~\ref{fig:architecture}. 
This architecture has the capability to 
 1) automatically rewrite all the  third party 
\emph{in-context} content of a Web application, 2) redirect the third party HTTP 
requests issued by the \emph{in-context} content, and 3) remove/disable known 
\emph{stateful} tracking mechanisms (see Section~\ref{sec:background}) 
for such third party content and requests. 4) It also rewrites and redirects \crosscontext\ requests so as to prevent website identification and communication with \incontext\ scripts.
%
% Figure~\ref{fig:architecture} presents the overall architecture we propose. 
% The main difference with traditional web application architecture is the introduction of two new components: a rewrite server and a middle party server.

%
Figure~\ref{fig:architecture} provides an overview of 
our web application architecture, that introduces two new 
components that are fully controlled by the website owner:

{\bf Rewrite Server (Section~\ref{sec:rewrite})} acts like a reverse 
proxy\footnote{\url{https://en.wikipedia.org/wiki/Reverse_proxy}} for the original web server. 
It rewrites the web pages in such a way that all the %\emph{controllable} 
third party requests %(which are possible to control, see Section~\ref{sec:problem}) 
are redirected through the \MPS\ before reaching the intended third party server. 

{\bf Middle Party Server (Section~\ref{sec:middle})} is at the core 
of our solution since it intercepts all browser third party requests, removes tracking, then forwards them 
to the intended third parties. When they reply, it also removes tracking information and forwards
the responses back to the browser. %all the third party requests to the third party server and removes stateful tracking. 
On one hand, it hides the third party destination from the browser, and therefore prevents the 
browser from attaching third party HTTP cookies to such requests. 
Because the browser will still attach some tracking information to the requests, such as \emph{ETag}, and 
\emph{Referer} headers, Middle Party Server will also remove this information when forwarding the 
requests to the third party. This prevents passive user identification for \incontext\ third party contents.

\begin{figure}[t]
    \centering
    \includegraphics[width=0.3\textwidth]{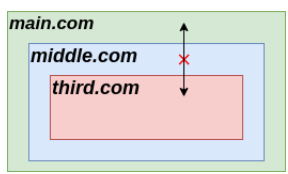}
    \caption{Prevent Combining in-context and cross-context tracking}
    \label{fig:crosscontext_rewriting}
    \end{figure}

On the other hand, the Middle Party Server prevents website identification for \crosscontext\ contents and communication with \incontext\ scripts.
This is done by placing the \crosscontext\ within another \crosscontext\ controlled by the Middle Party server as illustrated by Figure~\ref{fig:crosscontext_rewriting}.
For instance, if an iframe was to be embedded within a webpage, it is placed within another iframe that belongs to the Middle Party. 
%The iframe is loaded within this Middle Party iframe. 
The Middle Party then instructs the browser not to send \emph{Referer} header while loading the iframe,
which prevents passive and active website identification once it is loaded.
Since the iframe is nested within a iframe that belongs to Middle Party, this hides its reference to \incontext\ scripts (see Figure~\ref{fig:crosscontext_rewriting}).
Therefore, it is prevented from communicating with \incontext\ scripts in the main webpage.

%The Middle Party context the Rewrite server changes this URL
%To do so, is rewrites \crosscontext\ requests, so as to place the original \crosscontext not directly
%within the webpage, b
%both types of stateful tracking: 
%passive and active. For passive tracking (via HTTP headers), the server removes 
%tracking identifiers set by the third party, such as HTTP cookies and  Etag, 
%before forwarding the response back to the browser.
%For active tracking (via JavaScript code execution), \MPS\ 
%rewrites web pages before forwarding it back to the browser:
%it injects JavaScript code that prevents access to tracking APIs, 
%and cleans browser storages that are often used for tracking.
%and APIs used to exchange tracking information between \code{in-context} and \code{cross-context} cooperating third party 
%contents.
%The rewritting helps mitigate \code{cross-context} third party tracking.
 
% \subsection{Rewrite Serever}

\subsection{Rewrite Server}
\label{sec:rewrite}

The goal of the \RS\ is to rewrite the original content of the requested webpages
in such a way that all third party requests will be redirected 
to the \MPS. It consists of three main components: static HTML rewriter for HTML pages, 
static CSS rewriter and JavaScript injection component. Into each webpage, 
we inject a JavaScript code that insures that all the dynamically generated 
third party content is redirected to the \MPS. 

{\bf HTML and CSS Rewriter} rewrites the URLs of static third party contents 
embedded in original web pages and CSS files in order to redirect them to the \MPS. 
For example, the URL of a third-party script source \code{http://third.com/script.js}
is written so that it is instead fetched through the \MPS:
\code{http://middle.com/?src=http://third.com/script.js}. 
%The same process is followed for other static third party contents.

%\subsubsection{CSS File Rewriter} rewrites CSS files in order to ensure that all 
%the third party contents, like images and fonts, are redirected though the \MPS.
%%CSS files are a particular case, because they can further embed third party images and fonts. 
%%That's why we write them, just as we do with static third party contents in original web pages.
%We present an example of CSS rewriting in the Appendix.

{\bf JavaScript Injection.}
The \RS\ also injects a script in an original webpage, 
that controls APIs used to inject dynamic contents.
This injected script rewrites third party contents 
which are dynamically injected in webpages after 
they are rendered on the client-side.
  Table~\ref{tab:scripts_dynamic_apis} shows APIs that can be 
  used to dynamically inject third party content within a webpage 
  that we control using the injected script.  
  
\begin{table}[t]
  \centering
  \begin{tabular}{l|l}
  {\bf API} & {\bf Content}\\
  \hline
  \code{document.createElement} & inject contents from Table~\ref{tab:third_party_content}  \\
  \code{document.write} & any content\\
  \code{window.open} &  Web pages(popups) \\
  \emph{Image} & images \\
  \emph{XMLHttpRequest} & any data \\
  \emph{Fetch, Request} & any content \\
  \emph{Event Source} & stream data \\
  \emph{WebSocket} & websocket data\\
  %\emph{WebRTC} & P2P communication\\
  \hline
  \end{tabular}
\caption{Embedding Dynamic Third Party Contents}
\label{tab:scripts_dynamic_apis}
\SHORTEN
\end{table}

 A {\bf Content Security Policy (CSP)}~\cite{CSP2-W3C} is injected in the response header for 
 each webpage in order to prevent third parties from bypassing 
 the rewriting and redirection to the \MPS. 
 A CSP delivered with the webpage controls the resources of the 
 page. It allows to specify which resources are allowed to be loaded 
 and executed in the page. By limiting the resource origins to only 
 those from the \MPS\ and the website own domain, we prevent  third parties from bypassing 
 our redirection to the \MPS.

\label{sec:middle}
\subsection{Middle Party}
The main goal of the Middle Party is to proxy the requests and responses between browsers and third parties in order to remove tracking information exchanged between them. 
For \incontext\ contents, it removes any user recognition as well as website identification information. For \crosscontext\ contents, it takes care of preventing website identification
and communication with \incontext\ scripts.
%It removes tracking information in browser requests before fetching the related third party content on the original third party server. When it replies, it also removes
%any tracking information before forwarding the response back to the browser. In particular, it has a CSS Rewriter, for rewritten additional static third party contents embedded 
%in original third party CSS files. This is how it handles \incontext\ contents.
%Cross-context contents are handled to achieve two goals: prevent them from identifying the website and communicating with \incontext\ scripts. 

%They are reloaded within a \crosscontext controlled by the Middle Party. The goal is to prevent the third party content from identifying the website in which they are embedded. 

%Because of the SOP restrictions, we know exactly how stateful tracking can happen and how we can prevent them. 
{\bf In-Context Contents} are scripts, images, etc. (see Table~\ref{tab:third_party_content}).
%For example, 
Since a third party script from \code{http://third.com/script.js} is rewritten by the \RS\ 
to \code{http://middle.com/?src=http://third.com/script.js}, it is fetched 
through the \MPS.
When the middle party receives such a request URL from the browser, it takes the following steps.
{\bf Remove Tracking from request} that are set by the browser as HTTP headers. Among those headers are \emph{Cookie, Etag, If-Modified-Since, Cache-Control, Referer}. 
Next, it makes a request to the third party in order to get the content of the script \code{http://third.com/script.js}.
{\bf Remove Tracking from response} returned by the third party. The headers that the third party 
may send are \emph{Set-Cookie, Etag, Last-Modified, Cache-Control}. 
{\bf CSS Rewriter} rewrites the response if the content is a CSS file. Finally, the response is returned back to the browser.

{\bf Cross-context contents} are iframes, links,  popups, etc. (see Table~\ref{tab:third_party_content}).
For instance, a third party iframe from \code{http://third.com/page.html} is rewritten to \code{http:// middle.com/?emb=http://third.com/page.html}.
When the \MPS\ receives such a request URL from the browser, it takes the following actions:
{\bf URL Rewriting}: instead of fetching directly the content of \code{http://third.com/page.html}, the \MPS\ generates a content in which it puts the URL of the third party content as a hyperlink.
\code{<a href = "http://third.com/page.html" rel = "noreferrer noopener"></a>}.
The most important part of this content is in the \code{rel} attribute value. Therefore, \code{noreferrer noopener} instructs the browser not the send the \emph{Referer} header when the link \code{http://third.com/page.html}
is followed client-side.
{\bf JavaScript injection} module adds a script to the content so that the link gets automatically followed once the response is rendered by the browser.
Once the link is followed, the browser fetches the third party content directly on the third party server, without going through the Middle Party server anymore. 
Nonetheless, it does not include the \emph{Referer} header for identifying the website. Therefore, the \code{document.referrer} API also returns an empty string inside the iframe context. 
This prevents it from identifying the website.

The third party server response is placed within a new iframe nested within a context that belongs to the Middle Party, and not directly within the site webpage. This prevents \incontext\ scripts and 
the \crosscontext\ contents from exchanging tracking information as illustrated by Figure~\ref{fig:crosscontext_rewriting}.

\section{Implementation}
\label{sec:implementation}
We have implemented both the \RS\ and the \MPS\ %have been implemented 
as full Node.js~\cite{Node.js} web servers supporting \emph{HTTP(S)} protocols and web sockets.
Implementation details are available at \url{https://webstats.inria.fr/sstp/}.
%\NB{Not available}

\subsection{Rewrite Server}
%With the introduction of the rewrite server, the website owner has to manage two servers. All user requests should go through the rewrite server first. From a user perspective, the website should be
%accessible the same way. Let's consider that the website was accessible at \url{http://main.com}. From a developer point of 
{\bf Simple Forward}: requests that arrive to the Rewrite server are simply forwarded to the main server. 

%Technically, if the main server was accessed at \url{http://site.com:80}, it means that the original
%server was running on port 80. To support the rewrite server without having users change the way they access the website, on needs to run the rewrite server on the port 80, and run the original server on a different port,
%say 8080 for instance. The new port on which the original server runs is known to the rewrite server, so that it forwards requests on that port. 
%\NB{Text starting from ``Technically, if the main server'' until this comment 
%can be removed to gain space}

{\bf HTML Rewriter} is implemented with Jsdom HTML parser~\cite{Jsdom} 
and {\bf CSS Rewriter} using a CSS parser~\cite{CSSParser} for Node.js.
{\bf JavaScript injection} is done %at 
at the end of rewriting webpages. 
%\NB{during or after?}
The code script injected is available at \url{https://webstats.inria.fr/sstp/dynamic.js}.
%\NB{Not available}
{\bf CSP} set on webpages only whitelists the website own domain and the Middle Party. It also prevents third party plugins.
\begin{lstlisting}[]
Content-Security-Policy: default-src 'self' 'middle.com'; object-src 'self';
\end{lstlisting}
%\end{semiverbatim}

\subsection{Middle Party}
%\NB{Double-check the usage of $\backslash$ incontext $\backslash$}
{\bf In-Contexts Contents.}
{\bf Remove Tracking from requests} component removes tracking information from \incontext\ third party requests (See Section~\ref{sec:solution}). 
The requests are then forwarded to the original third party server, to fetch the third party content.
{\bf Remove Tracking from responses} : Tracking information that are set by third parties in the responses, are removed. %Also See 
See Section~\ref{sec:solution} for details about information that are removed.
{\bf CSS Rewriter}: as in the case of the Rewrite Server, this component is implemented using a a CSS parser~\cite{CSSParser} for Node.js for rewriting CSS files.

{\bf Cross-Context Contents.}
{\bf URL Rewriting} If the \crosscontext\ URL was \code{http://third.com/page.html}, this URL is rewritten to 
\begin{lstlisting}[]
<a href="http://third.com/page.html" rel="noreferrer noopener" target=""></a>.
\end{lstlisting}

{\bf JavaScript injection} : the content injected is as followed.
\begin{lstlisting}[]
var third_party = document.getElementsByTagName("a")[0];
if(window.top == window.self){
  third_party.target = "_blank";
  third_party.click();
  window.close();
}else{       
  var iframe = document.createElement("iframe");
      iframe.name = "iframetarget";
  document.body.appendChild(iframe);
  third_party.target = "iframetarget";
  third_party.click();
}
\end{lstlisting}
Both the rewritten URL and the injected script are returned as a response to the browser which renders it. The injected script gets executed within a context that belongs to the Middle Party.
If the original \crosscontext\ third party content was to be loaded inside an iframe, the injected script creates an iframe in which the original third party content is loaded. 
However, if it was to be opened inside a new tab, the injected script opens a new tab in which the third party content is loaded. 
In both cases, while the \crosscontext\ content is loaded, the browser does not sent the \emph{Referer} header. This makes the value of \code{document.referer} empty inside the \crosscontext\, preventing it 
from identifying the website.
Finally, since those \crosscontext\ are loaded by the injected script from a context that belongs to the Middle Party, \incontext\ scripts cannot communicate with the \crosscontext\ contents to
exchange tracking information. 

\subsection{Discussion and Limitations}
\label{sec:limitations}

%There are two main limitations in our implementation of which we are aware. 
Our approach suffers from the following limitations.
First, our implementation prevents \crosscontext\ and \incontext\ contents 
from communicating with each other using \code{postMessage} API. 
However,  \incontext\ third party script can identify the website a user visits 
via \code{document.location.href} API.
Then the script can include the website URL, say \code{http://main.com}, 
as a parameter of the URL of a third party iframe, for example 
 \code{http://third.com/page.html?ref=http://main.com} and 
 dynamically embed it in the webpage. 
 %, say an image, with the \code{src} field set to this URL.
%{\bf Passing website identification information are parameters of \crosscontext third party requests}: An \incontext third party script can identify the website, using for instance \code{document.location.href},
%and add this as a parameter to a \crosscontext third party URL, say \code{http://third.com/page.html?ref=http://main.com}. 
%
In our architecture, this URL is rewritten and routed to the Middle Party. Since, the \MPS\ does not inspect URL parameters, this information will reach the third party 
even though the \emph{Referer} is not sent with \crosscontext\ requests.

%\MPS\ then forwards the request to \code{third.com} but since we do not 
%analyse the content of  the URL, \code{third.com} will learn the website the 
%user visited.
%Since we did not
%inspect URLs for this kind of parameters, this will cause the information to the transmitted to the third party. Since we do not prevent user identification in \crosscontext, the user gets tracked.
%\NB{If we were removing passive user recognition in cross-context, 
%(removing HTTP cookies from the requests to third.com), then we wouldn't 
%have this limitation, right?}
Another limitation is that of dynamic CSS changes. For instance, changing the background image style of an element in the webpage is not captured by the dynamic rewriting script injected in webpages. 
Therefore, if the image was a third party image, the CSP will prevent it from loading.
%the case of 
%when CSS background images and fonts are dynamically injected 
%by third party scripts %once 
%after a webpage is rendered. 
%\NB{It's a limitation because we don't have control over dynamically 
%injected code after the rendering? If yes, rewrite the sentence accordingly}

{\bf Performance overhead}
There is a performance cost associated with the %introduction of a 
Rewrite Server.
Rewriting contents server-side and browser-side is also expensive in terms of performance. 
 \MPS\ may also lead to performance overhead especially for webpages with 
 numerous third party contents. 
%But the costly component is the Middle Party. In particular, if webpages that are rewritten have lots of third party contents.
%Nonetheless, one can use some 
We believe that server-side caching mechanisms may help to speed up responsiveness.
%requests through the Middle Party,
%

{\bf Extension to stateless tracking} 
Even though this work did not address stateless tracking, 
such as device fingerprinting,  
%there are some interesting side-effects worth noting.
our architecture already hides several fingerprintable device 
properties  and can be  extended to several others:
%\begin{itemize}
% \item 
1) The redirection to the Middle Party anonymizes the 
 real IP addresses of users; % to third parties.
2) % \item 
Some stateless tracking APIs such as \code{window.navigator}, 
 \code{window.screen}, and \code{HTMLCanvasElement} can be easily 
 removed or randomized from the context of the webpage to mitigate 
 \incontext\ fingerprinting. %stateless tracking.
%\end{itemize}

{\bf Possibility to blocking active user recognition in \crosscontext} 
With the prevalence of third party tracking on the web, we have shown the challenges that a developer will face towards mitigating that.
The sandbox attribute for iframes help prevent access to security-sensitive APIs. As tracking has become a hot concern, we suggest that 
similar mechanisms can help first party websites tackle third party tracking. The sandbox attribute can for instance be extended with 
specific values to tackle tracking. Nonetheless, the sandbox attribute can be used to prevent \crosscontext\ from some stateful tracking mechanisms~\cite{Sandbox}.
%has already interesting side-effects for mitigating some \crosscontext\ user identification. For instance, giving it the value \code{allow-scripts} prevents the use of browser storages,and plugins. 

\section{Evaluation and Case Study}
{\bf Demo website}
We have set up a demo website that embeds %consisting of 
a collection of %many 
third party contents, both \incontext\ and \crosscontext.
{ In-context} contents include images, HTML5 audio and video, and a Google Map, which further loads dynamic contents such as images, fonts, scripts, and CSS files;
a Youtube video as a \crosscontext\ content.
%The 
Our demo website is accessible at 
\url{http://sstp-rewriteproxy.inria.fr}. %, on port 80.
When we deployed the Rewrite Server on 
\url{http://sstp-rewriteproxy.inria.fr}, 
the original server has been moved to 
\url{http://sstp-rewriteproxy.inria.fr:8080}, so that it is no longer 
directly accessible to users. 
%the port 8080.
The Middle Party server runs at \url{http://sstp-middleparty.inria.fr}.

%In a normal browsing scenario, all those third party contents can participate in tracking. 
Originally, when all the third parties were simply included in the 
main webpage, they may have also been tracking the website users (see Figure~\ref{fig:tracking}).
After %With 
the deployment of our %this 
solution, we have been able to redirect all \incontext\ third party contents to the 
Middle Party. We have been able to prevent the website identification in the \crosscontext\ Youtube video. 
In the Appendix, we show a screenshot of requests redirection to the \MPS. 
%The demo is available online, and one can check it out.

{\bf Real websites}
%We wanted to also evaluate our implementation on real case websites. However, 
%since 
Since we did not have access to a real websites, we cannot install a 
\RS\ and  to evaluate our solution. We therefore 
%we have set up the following scenario.
%\begin{itemize}
% \item We 
implemented a browser proxy %. It is 
based on a Node.js proxy~\cite{ProxyNode.js}, and included 
 %\item We implemented 
 all the logic of the \RS\ % rewrite server 
 within the proxy. The proxy is running at \url{http://sstp-rewriteproxy.inria.fr:5555}. 
%\end{itemize}

We then evaluated the solution on different kinds of websites: % including 
a news website \url{http://www.bbc.com},  
an entertainment website \url{http://www.imdb.com}, %(an entertainement website), 
and a shopping website \url{http://verbaudet.fr}. %, a shopping website. 
%They all 
All three websites
load content from various third party domains. 
In all %those 
websites, we %blindly 
rewrote all third party contents through the proxy (acts as \RS) and the \MPS\ removed tracking information.
% 
%In a normal deployment scenario, one would consider not rewritting tracking content since those contents are not
%the target of this solution. In particular, since we remove the \emph{Referer} header in HTTP requests, we may break some contents features if this header is ever required by the third party to customize
%responses. This also hardens the evaluation of our solution. 
Visually, we did not notice any change in the behaviors of the websites. 
We also interacted with them in a standard way (clicking on links on a news 
website, choosing products and putting them in the basket on the shopping website) 
and all the main functionalities of the websites was preserved.
%\NB{Francis, rewrite above it not true} 

Overall, %though, those 
these evaluation scenarios have helped us improve the solution, 
especially rewriting dynamically injected third party content. 
We believe that this implementation will even get better in the future 
when we convince to deploy it for some real websites. 
\section{Related Work}
Many studies have demonstrated that third party tracking is very prevalent on the web today as well as the underlying
tracking technologies~\cite{Engl-Nara-16-CCS, Roes-etal-12-NSDI, Maye-Mitc-12-SP, Kris-Will-09-WWW}. Lerner et al.~\cite{Lern-etal-16-USENIX} dusted the story of this practice for a period of twenty years.
Trackers have been categorized according to either their business relationships with websites~\cite{Maye-Mitc-12-SP}, their prominence~\cite{Kris-Will-09-WWW, Engl-Nara-16-CCS} or the user browsing profile that they can build~\cite{Roes-etal-12-NSDI}.
Mayer and Mitchell~\cite{Maye-Mitc-12-SP} grouped tracking mechanisms in two categories called statefull (cookie-based and super-cookies) and stateless (fingerprinting).
It is rather intuitive to convince ourselves about the effectiveness of a statefull tracking, since the latter is based on unique identifiers that
are set in users browsers. Nonetheless, the efficacy of stateless mechanisms has been extensively demonstrated. Since the pioneer work of Eckersley~\cite{Eckersley-10-PETS},
new fingerprinting methods have been revealed in the literature~\cite{Star-Niki-17-WWW, Cao-etal-17-NDSS, Engl-Nara-16-CCS, Acar-etal-13-CCS, Boda-etal-11-Nordec, Abgr-etal-12-CoRR, Take-etal-15-BWCCA, Niki-etal-13-SP, Acar-etal-14-EEJND}.
A classification of fingerprinting techniques is provided in~\cite{Upat-etal-15-NTMS}.
Those studies have contributed to raising public awareness of tracking privacy threats. Mayer and Mitchell~\cite{Maye-Mitc-12-SP} have shown that users are very sensitive to their online privacy, thus hostile to third party
tracking. Englehardt et al.~\cite{Engl-etal-15-WWW} have demonstrated that tracking can be used for surveillance purposes. The success of anti-tracking defenses is yet another illustration of users concern regarding tracking~\cite{Merz-etal-17-EuroSnP}. 

There are many defenses that try to protect users against third party tracking. First, major browser vendors do natively provide mechanisms for users to block third party cookies, browse in private mode.
More and more privacy-browsers even take a step further, putting privacy as a design and implementation principle. Examples of such browsers are the Tor Browser~\cite{TorBrowser}, TrackingFree Browser~\cite{Pan-etal-15-NDSS}
or Blink~\cite{Lape-etal-16-SP}.
But the most popular defenses are by far browser extensions. Being tightly integrated to browsers, they provide additional privacy features that are not natively implemented in browsers. Well known 
extensions for privacy are Disconnect~\cite{Disconnect}, Ghostery~\cite{Ghostery}, AdBlock~\cite{AdBlock}, ShareMeNot~\cite{ Roes-etal-12-NSDI}, which is now part of PrivacyBadger~\cite{PrivacyBadger},
MyTrackingChoices~\cite{Acha-etal-16-CoRR}, MyAdChoices~\cite{Parr-etal-16-CoRR}. Merzdovnik et al.~\cite{Merz-etal-17-EuroSnP} provide a large-scale study of anti-tracking defenses.
Well known trackers such as advertisers, which businesses hugely depend on tracking, have also been taking steps towards limiting their tracking capabilities~\cite{Maye-Mitc-12-SP}. 
The W3C is pushing forward the Do Not Tracking standard~\cite{DNT-TCS-W3C, DNT-TPE-W3C} for users to easily express their tracking preferences so that trackers may comply with them.
%Third party contents are valuable to website owners and very prevalent among them~\cite{Niki-etal-12-CCS}. With the privacy risk of embedding them in websites, it is important to find a tradeoff between
%embedding third party contents and preserving users privacy. Current tracking defenses, have mostly focused on trackers only even though any third party content can participate in tracking. 
To the best of our knowledge, we are the first to investigate how a website owner can embed third party content while preventing them from accidentally tracking users. The idea of proxying requests
within a webpage is inspired by web service workers API~\cite{ServiceWorker}, though the latter is still a working draft which is being tested in 
Mozilla Firefox and Google Chrome.

\section{Conclusions}
Most of the previous research analysed %Much of work has been done regarding 
third party tracking mechanisms, and how to block tracking from a user perspective. 
%, especially about the mechanisms, and also
%more and more defenses for users. 
In this work, we classified third party tracking capabilities %regarding 
from a website developer perspective. 
%In particular, we 
We proposed a new architecture for %them 
website developers that allows to embed third party
contents while preserving users privacy. 
We implemented our solution, and evaluated it on real websites to %for
mitigate stateful tracking. 
%We believe that this architecture can be easily extended to handle 
%stateless tracking. % mitigations.
%We hope that this architecture will serve as a basis for building future privacy-preserving websites, especially
%in the era of tracking. 
%\newpage

\newpage
\bibliographystyle{abbrv}
\bibliography{main.bib}

\begin{thebibliography}{10}

\bibitem{AdBlock}
{AdBlock - Block Ads - Browse Safe}.
\newblock \url{https://getadblock.com/}.

\bibitem{Context}
{Browsing Contexts}.
\newblock \url{https://www.w3.org/TR/html51/browsers.html}.

\bibitem{CSS}
{Cascading Style Sheets}.
\newblock \url{https://www.w3.org/Style/CSS/}.

\bibitem{CORS}
Cross-origin-resource sharing.
\newblock
  \url{https://developer.mozilla.org/en-US/docs/Web/HTTP/Access_control_CORS}.

\bibitem{CSSParser}
{CSS Parser for Node.js}.
\newblock \url{https://github.com/reworkcss/css}.

\bibitem{Disconnect}
{Disconnect}.
\newblock \url{https://disconnect.me/}.

\bibitem{Ghostery}
{Ghostery}.
\newblock \url{https://www.ghostery.com/}.

\bibitem{Jsdom}
{HTML Parser for Node.js}.
\newblock \url{https://github.com/tmpvar/jsdom}.

\bibitem{Sandbox}
{Iframe Sandbox Attribute}.
\newblock
  \url{https://www.w3.org/TR/2011/WD-html5-20110525/the-iframe-element.html#attr-iframe-sandbox}.

\bibitem{Node.js}
{Node.js}.
\newblock \url{https://nodejs.org/en/}.

\bibitem{ProxyNode.js}
{Node.js Proxy}.
\newblock
  \url{https://newspaint.wordpress.com/2012/11/05/node-js-http-and-https-proxy}.

\bibitem{PostMessage}
{PostMessage - Cross-Origin Iframe Secure Communication}.
\newblock
  \url{https://developer.mozilla.org/en-US/docs/Web/API/Window/postMessage}.

\bibitem{PrivacyBadger}
{Privacy Badger - Electronic Frontier Foundation}.
\newblock \url{https://www.eff.org/fr/privacybadger}.

\bibitem{Same-Origin-Policy}
{Same Origin Policy}.
\newblock \url{https://www.w3.org/Security/wiki/Same_Origin_Policy}.

\bibitem{ServiceWorker}
{Service Worker API}.
\newblock
  \url{https://developer.mozilla.org/en-US/docs/Web/API/Service_Worker_API}.

\bibitem{TorBrowser}
{The Design and Implementation of the Tor Browser [Draft]}.
\newblock \url{https://www.torproject.org/projects/torbrowser/design/}.

\bibitem{Abgr-etal-12-CoRR}
E.~Abgrall, Y.~L. Traon, M.~Monperrus, S.~Gombault, M.~Heiderich, and
  A.~Ribault.
\newblock {XSS-FP:} browser fingerprinting using {HTML} parser quirks.
\newblock {\em CoRR}, abs/1211.4812, 2012.

\bibitem{Acar-etal-14-EEJND}
G.~Acar, C.~Eubank, S.~Englehardt, M.~Ju{\'{a}}rez, A.~Narayanan, and
  C.~D{\'{\i}}az.
\newblock The web never forgets: Persistent tracking mechanisms in the wild.
\newblock In G.~Ahn, M.~Yung, and N.~Li, editors, {\em Proceedings of the 2014
  {ACM} {SIGSAC} Conference on Computer and Communications Security,
  Scottsdale, AZ, USA, November 3-7, 2014}, pages 674--689. {ACM}, 2014.

\bibitem{Acar-etal-13-CCS}
G.~Acar, M.~Ju{\'{a}}rez, N.~Nikiforakis, C.~D{\'{\i}}az, S.~F. G{\"{u}}rses,
  F.~Piessens, and B.~Preneel.
\newblock Fpdetective: dusting the web for fingerprinters.
\newblock In A.~Sadeghi, V.~D. Gligor, and M.~Yung, editors, {\em 2013 {ACM}
  {SIGSAC} Conference on Computer and Communications Security, CCS'13, Berlin,
  Germany, November 4-8, 2013}, pages 1129--1140. {ACM}, 2013.

\bibitem{Acha-etal-16-CoRR}
J.~P. Achara, J.~Parra{-}Arnau, and C.~Castelluccia.
\newblock Mytrackingchoices: Pacifying the ad-block war by enforcing user
  privacy preferences.
\newblock {\em CoRR}, abs/1604.04495, 2016.

\bibitem{Boda-etal-11-Nordec}
K.~Boda, {\'{A}}.~M. F{\"{o}}ldes, G.~G. Guly{\'{a}}s, and S.~Imre.
\newblock User tracking on the web via cross-browser fingerprinting.
\newblock In P.~Laud, editor, {\em Information Security Technology for
  Applications - 16th Nordic Conference on Secure {IT} Systems, NordSec 2011,
  Tallinn, Estonia, October 26-28, 2011, Revised Selected Papers}, volume 7161
  of {\em Lecture Notes in Computer Science}, pages 31--46. Springer, 2011.

\bibitem{Cao-etal-17-NDSS}
Y.~Cao, S.~Li, and E.~Wijmans.
\newblock (cross-)browser fingerprinting via os and hardware level features.
\newblock In {\em 24th Annual Network and Distributed System Security
  Symposium, {NDSS} 2017, San Diego, California, USA, 26 February - 1 March,
  2017}, 2017.
\newblock To Appear.

\bibitem{DNT-TCS-W3C}
N.~Doty.
\newblock {Tracking Compliance and Scope}, 2016.
\newblock \url{https://www.w3.org/TR/tracking-compliance/}.

\bibitem{Eckersley-10-PETS}
P.~Eckersley.
\newblock How unique is your web browser?
\newblock In M.~J. Atallah and N.~J. Hopper, editors, {\em Privacy Enhancing
  Technologies, 10th International Symposium, {PETS} 2010, Berlin, Germany,
  July 21-23, 2010. Proceedings}, volume 6205 of {\em Lecture Notes in Computer
  Science}, pages 1--18. Springer, 2010.

\bibitem{Engl-Nara-16-CCS}
S.~Englehardt and A.~Narayanan.
\newblock Online tracking: {A} 1-million-site measurement and analysis.
\newblock In E.~R. Weippl, S.~Katzenbeisser, C.~Kruegel, A.~C. Myers, and
  S.~Halevi, editors, {\em Proceedings of the 2016 {ACM} {SIGSAC} Conference on
  Computer and Communications Security, Vienna, Austria, October 24-28, 2016},
  pages 1388--1401. {ACM}, 2016.

\bibitem{Engl-etal-15-WWW}
S.~Englehardt, D.~Reisman, C.~Eubank, P.~Zimmerman, J.~Mayer, A.~Narayanan, and
  E.~W. Felten.
\newblock Cookies that give you away: The surveillance implications of web
  tracking.
\newblock In A.~Gangemi, S.~Leonardi, and A.~Panconesi, editors, {\em
  Proceedings of the 24th International Conference on World Wide Web, {WWW}
  2015, Florence, Italy, May 18-22, 2015}, pages 289--299. {ACM}, 2015.

\bibitem{DNT-TPE-W3C}
R.~T. Fielding.
\newblock {Tracking Preference Expression (DNT)}, 2015.
\newblock \url{https://www.w3.org/TR/tracking-dnt/}.

\bibitem{Kris-Will-09-WWW}
B.~Krishnamurthy and C.~E. Wills.
\newblock Privacy diffusion on the web: a longitudinal perspective.
\newblock In J.~Quemada, G.~Le{\'{o}}n, Y.~S. Maarek, and W.~Nejdl, editors,
  {\em Proceedings of the 18th International Conference on World Wide Web,
  {WWW} 2009, Madrid, Spain, April 20-24, 2009}, pages 541--550. {ACM}, 2009.

\bibitem{Lape-etal-16-SP}
P.~Laperdrix, W.~Rudametkin, and B.~Baudry.
\newblock Beauty and the beast: Diverting modern web browsers to build unique
  browser fingerprints.
\newblock In {\em {IEEE} Symposium on Security and Privacy, {SP} 2016, San
  Jose, CA, USA, May 22-26, 2016}, pages 878--894. {IEEE} Computer Society,
  2016.

\bibitem{Lern-etal-16-USENIX}
A.~Lerner, A.~K. Simpson, T.~Kohno, and F.~Roesner.
\newblock Internet jones and the raiders of the lost trackers: An
  archaeological study of web tracking from 1996 to 2016.
\newblock In {\em 25th USENIX Security Symposium (USENIX Security 16)}, Austin,
  TX, 2016. USENIX Association.

\bibitem{Maye-Mitc-12-SP}
J.~R. Mayer and J.~C. Mitchell.
\newblock Third-party web tracking: Policy and technology.
\newblock In {\em {IEEE} Symposium on Security and Privacy, {SP} 2012, 21-23
  May 2012, San Francisco, California, {USA}}, pages 413--427. {IEEE} Computer
  Society, 2012.

\bibitem{Merz-etal-17-EuroSnP}
G.~Merzdovnik, M.~Huber, D.~Buhov, N.~Nikiforakis, S.~Neuner, M.~Schmiedecker,
  and E.~Weippl.
\newblock Block me if you can: A large-scale study of tracker-blocking tools.
\newblock In {\em 2nd IEEE European Symposium on Security and Privacy}, Paris,
  France, 2017.
\newblock To appear.

\bibitem{Niki-etal-13-SP}
N.~Nikiforakis, A.~Kapravelos, W.~Joosen, C.~Kruegel, F.~Piessens, and
  G.~Vigna.
\newblock Cookieless monster: Exploring the ecosystem of web-based device
  fingerprinting.
\newblock In {\em 2013 {IEEE} Symposium on Security and Privacy, {SP} 2013,
  Berkeley, CA, USA, May 19-22, 2013}, pages 541--555. {IEEE} Computer Society,
  2013.

\bibitem{Pan-etal-15-NDSS}
X.~Pan, Y.~Cao, and Y.~Chen.
\newblock I do not know what you visited last summer: Protecting users from
  stateful third-party web tracking with trackingfree browser.
\newblock In {\em 22nd Annual Network and Distributed System Security
  Symposium, {NDSS} 2015, San Diego, California, USA, February 8-11, 2015}. The
  Internet Society, 2015.

\bibitem{Parr-etal-16-CoRR}
J.~Parra{-}Arnau, J.~P. Achara, and C.~Castelluccia.
\newblock Myadchoices: Bringing transparency and control to online advertising.
\newblock {\em CoRR}, abs/1602.02046, 2016.

\bibitem{Roes-etal-12-NSDI}
F.~Roesner, T.~Kohno, and D.~Wetherall.
\newblock Detecting and defending against third-party tracking on the web.
\newblock In S.~D. Gribble and D.~Katabi, editors, {\em Proceedings of the 9th
  {USENIX} Symposium on Networked Systems Design and Implementation, {NSDI}
  2012, San Jose, CA, USA, April 25-27, 2012}, pages 155--168. {USENIX}
  Association, 2012.

\bibitem{Solt-etal-10-AAAI}
A.~Soltani, S.~Canty, Q.~Mayo, L.~Thomas, and C.~J. Hoofnagle.
\newblock Flash cookies and privacy.
\newblock In {\em AAAI spring symposium: intelligent information privacy
  management}, pages 158--163, 2010.

\bibitem{Star-Niki-17-WWW}
O.~Starov and N.~Nikiforakis.
\newblock Extended tracking powers: Measuring the privacy diffusion enabled by
  browser extensions.
\newblock In {\em Proceedings of the 26th International Conference on World
  Wide Web, {WWW} 2017, Perth, Australia, April 3 - 7, 2017}, 2017.
\newblock To Appear.

\bibitem{Take-etal-15-BWCCA}
N.~Takei, T.~Saito, K.~Takasu, and T.~Yamada.
\newblock Web browser fingerprinting using only cascading style sheets.
\newblock In L.~Barolli, F.~Xhafa, M.~R. Ogiela, and L.~Ogiela, editors, {\em
  10th International Conference on Broadband and Wireless Computing,
  Communication and Applications, {BWCCA} 2015, Krakow, Poland, November 4-6,
  2015}, pages 57--63. {IEEE} Computer Society, 2015.

\bibitem{Upat-etal-15-NTMS}
R.~Upathilake, Y.~Li, and A.~Matrawy.
\newblock A classification of web browser fingerprinting techniques.
\newblock In M.~Badra, A.~Boukerche, and P.~Urien, editors, {\em 7th
  International Conference on New Technologies, Mobility and Security, {NTMS}
  2015, Paris, France, July 27-29, 2015}, pages 1--5. {IEEE}, 2015.

\bibitem{CSP2-W3C}
M.~West, A.~Barth, and D.~Veditz.
\newblock {Content Security Policy Level 2. W3C Candidate Recommendation},
  2015.

\end{thebibliography}

\section*{Appendix}
Screenshot of the demo website map console.
\iffalse
\begin{figure}[!tp]
\includegraphics[width=1\textwidth]{images/screen1}
\caption{Screen 1}
\label{fig:screen1}
\end{figure}
\fi
\begin{figure}[!htp]
\includegraphics[width=1\textwidth]{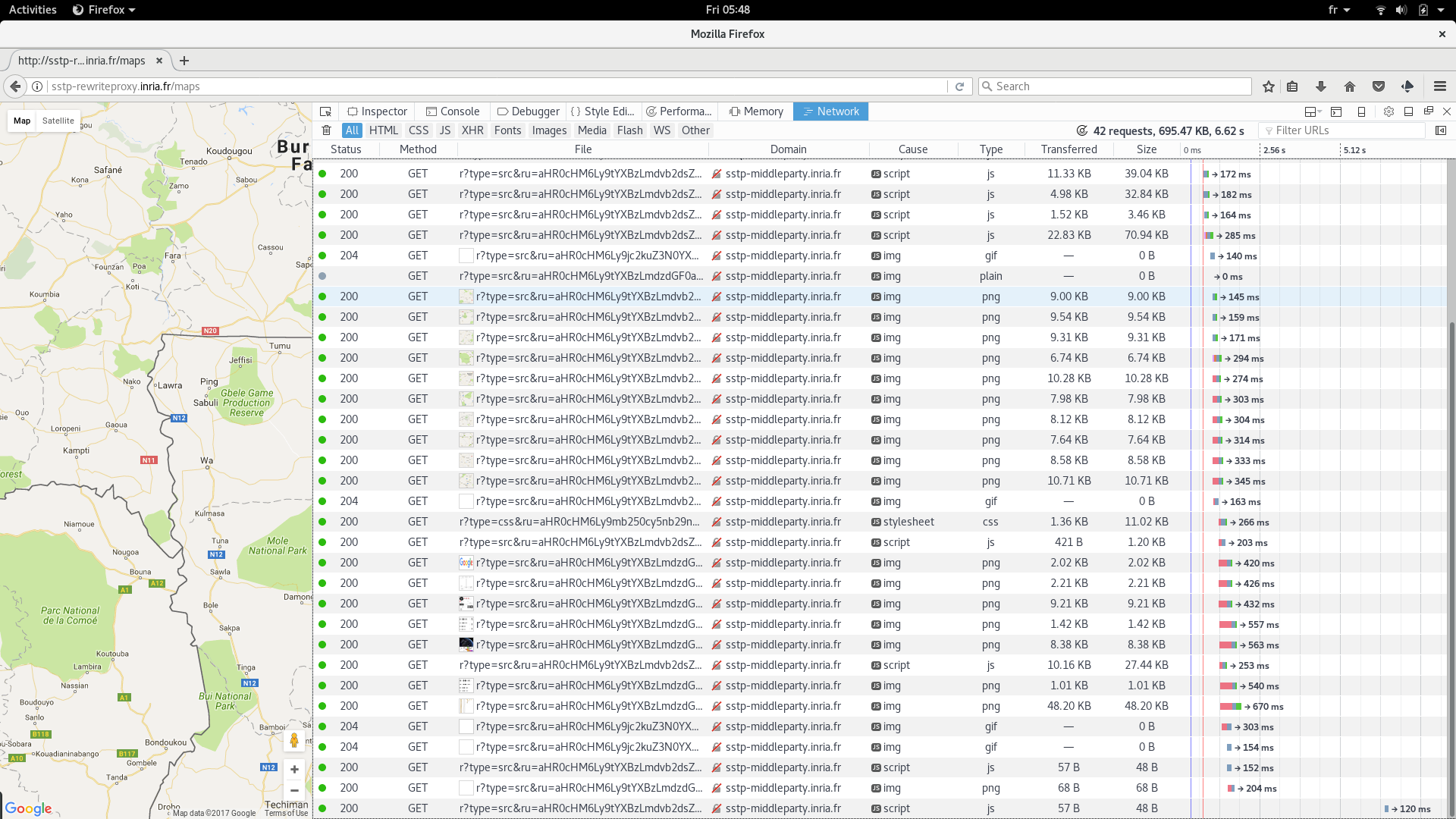}
\caption{Screenshot of the Browser console}
\label{fig:screen2}
\end{figure}

\end{document}